\documentclass[letter,twocolumn]{IEEEtranTCOM}
\usepackage{longtable}
\usepackage{amsmath}
\usepackage{cancel}
\usepackage{amssymb}
\usepackage{graphicx}
\usepackage{longtable}
\usepackage{multirow}
\usepackage{array}
\usepackage{balance}
\usepackage[labelsep=period]{caption}
\usepackage{setspace}
\usepackage{float}
\usepackage{subfig}
\usepackage{lettrine}
\usepackage[noadjust]{cite}
\usepackage{cleveref}
\usepackage[english]{babel}
\usepackage{tikz}
\usepackage[letterpaper, left=0.625in, right=0.625in, bottom=1in, top=0.75in]{geometry}
\usepackage[figurename=Fig.]{caption}
\renewcommand{\figurename}{Fig.}

\usepackage{mathtools,amssymb,lipsum}

\begin{document}
\title{Recent Advancements in Defected Ground Structure Based Near-Field Wireless Power Transfer Systems}
\author{Kassen Dautov, 
		Mohammad Hashmi, \IEEEmembership{Senior Member,~IEEE,}
		Galymzhan Nauryzbayev, \IEEEmembership{Member,~IEEE,} \\
		and M. Nasimuddin, \IEEEmembership{Senior Member,~IEEE} 
		\thanks{This work was supported by the Nazarbayev University FDRG Program under Grants SOE2019005 (110119FD4515) and 240919FD3935.}
		\thanks{K. Dautov, M. Hashmi, and G. Nauryzbayev are with the Department of Electrical and Computer Engineering, School of Engineering and Digital Sciences, Nazarbayev University, Nur-Sultan, 010000, Kazakhstan (e-mail: \{kassen.dautov, mohammad.hashmi, galymzhan.nauryzbayev\}@nu.edu.kz).}
		\thanks{M. Nasimuddin is with the Institute for Infocomm Research, A-STAR, 138632, Singapore (e-mail: nasimuddin@i2r.a-star.edu.sg).}
}
\maketitle
\thispagestyle{empty}
\begin{abstract}
\boldmath
The defected ground structure (DGS) technique enables miniaturization of the resonator which leads to the development of the compact near-field wireless power transfer (WPT) systems. In general, numerous challenges are inherent in the design of the DGS-based WPT systems and, hence, appropriate trade-offs for achieving optimal performance are required. Furthermore, the design advancements have led to the development of the DGS-based multi-band WPT systems to fulfill the needs of simultaneous data and power transfer. The innovations in the DGS-based WPT systems have also resulted in the definition of more commonly used figures-of-merit for the benchmarking of various performance metrics. The literature is replete with the design schemes to address one or more associated design challenges and successful WPT system realizations with enhanced performance. With this in mind, this paper touches upon the DGS-based WPTs developments and presents a concise report on the current state-of-the-art and future directions.
\end{abstract}


\IEEEpeerreviewmaketitle

\section{Introduction}
\label{Intro}
\PARstart{N}{owadays} replacing wires by applying wireless power transfer (WPT) systems has become very attractive since the latter enables the charging of multiple electronic devices simultaneously \cite{Ng} and powering the sensors with very low power consumption \cite{Jonah-1,Kung-2}. Transmission of electrical power through free space using radio waves, \textit{i.e.}, essentially entails the transmission of electrical energy without employing any physical link. The concept of WPT can be traced back to $1888$ when Heinrich Hertz demonstrated, for the first time, wireless transmission of power \cite{Hertz-1,Brown-1}. However, it is also imperative to mention that the modern world considers Nikola Tesla as the father of WPT as he was the first to experiment and document his work on WPT way back in $1899$  \cite{Tesla-1,Tesla-2,Shinohara-1}.

\begin{figure}[t!]
	\centering
	\includegraphics[width=1\linewidth]{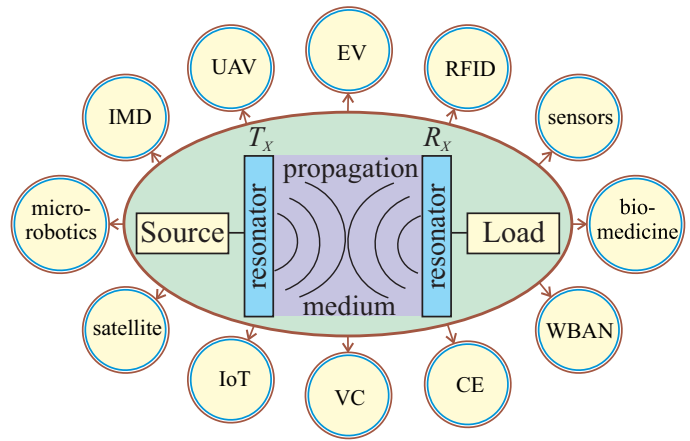}
	\caption{The potential applications of the WPT systems.}
	\label{fig: WPT system}
\end{figure}

In general, it is feasible to divide WPTs into two categories, namely, far-field WPT~\cite{Belo-1} and near-field WPT~\cite{Garnica-1,Kung-1}. The far-field WPTs, which are radiative in nature, find usefulness in applications such as satellite communications \cite{Spadden-1}, unmanned aerial vehicles (UAVs) \cite{Xu-1}, and vehicular communication (VC) \cite{Ghotbi-1}. On the other hand,  the near-field WPTs (\textit{i.e.,} non-radiative) require a short range and are useful for applications including but not limited to implantable microelectronic devices (IMDs)~\cite{Kiani-1,Zargham-1,Wang-2}, radio-frequency identifiers (RFIDs) \cite{Chen-1}, drug delivery systems \cite{Tahar-1}, micro-robotics \cite{Cannon-1}, consumer electronics (CE) \cite{Jolani-2,Jolani-1}, wireless body area network (WBAN) \cite{Mohamed-1,Rano-2,Ling-1}, electrical vehicles (EVs) charging \cite{Huh-1,Hsieh-1}, and sensors employed in biomedical domain~\cite{Jia-1,Mei-1,Freeman-1}. A generic schematic of WPT and its potential applications are depicted in Figure~\ref{fig: WPT system}. In this context, it is pertinent to mention that numerous WPT techniques, which revolve around concepts such as inductive coupling \cite{Jow-1}, magnetic resonance \cite{Zhang-1}, meta-surfaces~\cite{R1,R2,Li-1}, capacitive coupling \cite{Liu-3}, lasers \cite{R3}, \textit{etc}., have been proposed to facilitate both near- and far-field WPT applications. 

\begin{table*}[t!]
	\centering
	\footnotesize
	\caption{The parametric comparison of the DGS- and coil-based near-field WPT techniques.}
	\label{tab: comparison}
	\begin{tabular}{|l|c|c|}
		\hline
		\multicolumn{1}{|c|}{\textbf{Parameters}} & \textbf{DGS} & \textbf{Coil (incl. planarized)} \\ \hline
		Operating frequency\cite{Jolani-1,Falavarjani-1,My-2,Tahar-1} & high ($>100$~MHz) & low ($<100$~MHz) \\ \hline
		Power transfer efficiency \cite{Falavarjani-1,Barakat-1,Tahar-1,Barakat-3,Chalise-1} & moderate ($60-75\%$) & low ($<50\%$ at low frequencies)\\ \hline
		Power transfer distance \cite{My-2,Kurs-1,Jolani-2,Hekal-1} & moderate & high \\ \hline
		Implementation \cite{Jolani-1,Falavarjani-1,Hekal-1} & easy & complicated \\ \hline
		Equivalent circuit \cite{Karmakar-1,Cannon-1,Hekal-5} & complicated (quasi-static) & easy \\ \hline
		Multi-band operation \cite{My-2,Barakat-3,Tahar-1,Kupreyev-1} & $+$ & $-$ \\ \hline
		Frequency split phenomenon \cite{Zhang-1,My-2,Verma-2,Tahar-2,Sharaf-2} & $+$ & $+$ \\ \hline
		Robustness \cite{Hekal-1,Hekal-2} & $+$ & $-$ \\ \hline
		High $Q$ realization \cite{Jolani-1,Hekal-4,Verma-1} & $+$ & $+$ (when planarized) \\ \hline
		Cost \cite{Falavarjani-1,Hekal-2} & $-$ & $+$ \\ \hline
		Compactness \cite{Tahar-1,My-2,Hekal-2,Hekal-5} & $+$ & $-$ \\ \hline
		Ease of design \cite{Verma-1,Chen-1} & $+$ & $-$ \\ \hline
		Misalignment sensitivity \cite{Liu-2,Barakat-1,Hekal-2} & $+$ & $+$ \\ \hline
	\end{tabular}
\end{table*}

Within the scope of near-field WPTs, three distinct coupling-based techniques, namely, capacitive \cite{Liou-1}, inductive~\cite{Riehl-1}, and magnetic resonance~\cite{Kurs-1}, have been widely explored. Out of these, the capacitive one possesses a smaller power transfer range as compared to the other approaches and, therefore, finds limited usefulness in practical applications~\cite{Lu-1}. On the other hand, the inductive and resonant coupling-based WPTs, with a relatively longer power transmission distance, are built around metallic coils~\cite{Kim-1}. There have been a number of advancements in such techniques, considering that the evolution in technology enabled the design of printed coils and spirals using microstrip lines (MLs) \cite{Wang-1,Falavarjani-1}. Several design reports related to the resonance coupling-based WPT systems focus power on certain carrier frequency and, as a consequence, the efficiency of the system improves. Therefore, such WPTs exhibit a better performance as compared to the systems based on the inductive coupling~\cite{Barman-1,Hui-1}. In general, the realization of the resonance coupling requires a resonator~\cite{Shinohara-2}, and a defected ground structure (DGS) technique can aid in the development of the needed resonators that possess quasi-lumped resonant circuit behavior~\cite{Hekal-2,Hekal-4}. In this regard, it is imperative to mention that resonance coupling-based near-field WPTs can be achieved using either the coil (including planarized ones) or DGS. Therefore, keeping the specifications and other design goals in perspective, these approaches are compared in Table~\ref{tab: comparison}.

A simple literature search related to the near-field WPT systems using coils (including the printed ones) conveys that a number of innovative design techniques, analysis methods, and performance-enhancing approaches have been reported \cite{Zhang-1,Lee-1,Liu-2,Mastri-1,Raju-1,Ahn-2,Ghovanloo-1,Li-2}. For example, the WPT performance worsens in the presence of lateral and angular misalignment between the transmitting and receiving coils, and this issue was addressed by a new design configuration~\cite{Liu-2}. There are often mutual coupling issues between the transmitter ($T_{X}$) and receiver ($R_{X}$) and that can be mitigated by the incorporation of a switchable circuit with capacitors in designed WPT~\cite{Lee-1}. It has been established that the mutual inductance ($M$) and coupling between resonators significantly impact the performance of coil-based WPT systems and~\cite{Raju-1} proposed an analytical solution that can in advance predict $M$ between the planar coils. Sometimes, there may be issues due to over-coupling as it creates a split in the resonant frequency and, hence, in the performance of WPT~\cite{Zhang-1}, and an analytical technique to address this concern related to the frequency splitting phenomenon was reported in~\cite{Mastri-1}. It is pertinent to mention that coupling-based WPT techniques rely on a high-quality factor ($Q$) of resonators to enhance the performance. This was aptly demonstrated through a very intuitive design using a double-sided substrate integrated suspended line technology to enhance the $Q$ factor of the conventional printed spiral inductors~\cite{Li-2}. Some of the reported spiral-based WPT systems were application-specific, \textit{e.g.}, a high-efficiency WPT for IMDs by considering various standardized frequencies~\cite{Ahn-2} and different signal carriers~\cite{Ghovanloo-1}. A brief survey on the coil-based WPT systems reveals that these systems are promising and can transfer power to relatively longer distances. However, they are limited in utility at higher frequencies in the radio waves spectrum owing to large circuit dimensions \cite{Cannon-1,Kurs-1}, fabrication complexity \cite{Hekal-1},  and the bulkiness of the system \cite{Hekal-2}. Planarizing the coils mitigates such challenges to some extent but, still, these do not fulfill the needs of applications requiring miniaturization and simpler design with very high efficiency in lieu of the reduced power transmission range \cite{Lim-7,RamRakhyani-1,Ahn-3}.

The coupling-based WPT systems and techniques developed using DGS resonators have been getting tremendous attention in the past few years. One of the key factors contributing to this traction is the ability of DGS to provide definite benefits in terms of the miniaturization and compactness of the WPT systems \cite{Lou-1}. Furthermore, other benefits of this technique over the coil-based approach include the ease of design and implementation \cite{Verma-1}, robustness and reliability \cite{Hekal-1}, cost-effectiveness and relatively lower losses \cite{Hekal-2}, and more freedom in the realization of high $Q$ by controlling the geometry of DGS shapes \cite{Chalise-1}. The major challenges encountered by the DGS-based WPT systems are relatively smaller power transfer distance \cite{My-2,Verma-2} and complicated quasi-static equivalent circuit of DGS shapes~\cite{Hekal-1,Karmakar-1}. In this instance, it is worth mentioning that numerous challenges in the design of such systems occur due to the trade-offs among the power transmission range, efficiency, size, and design complexity. However, any comprehensive yet succinct information about these aspects is unavailable. Therefore, this paper aims to fill this gap by providing an extensive resource about the challenges, design techniques to overcome them, performance enhancement approaches, and research directions related to the DGS-based WPT systems.    

The organization of the rest of this article is as follows. Section \ref{sec: Literature Review} elaborates on the state-of-the-art of the DGS-based WPT systems. Section \ref{sec: WPT Challenges} introduces the major challenges of DGS-based WPTs as well as the features of the DGS technique, analyses of various DGS shapes, and illustration of the main design steps. Section \ref{sec: Realized WPTs} presents some successfully implemented DGS-based WPT systems in recent years. Finally, the potential research directions of the DGS-based WPT systems are elaborated in Section \ref{Sec: Future Works}, while Section \ref{Conclusion} concludes the paper.


\section{Recent Advancements of DGS-based WPT\lowercase{s}}
\label{sec: Literature Review}
\begin{figure}[t!]
	\centering
	\includegraphics[width=1\linewidth]{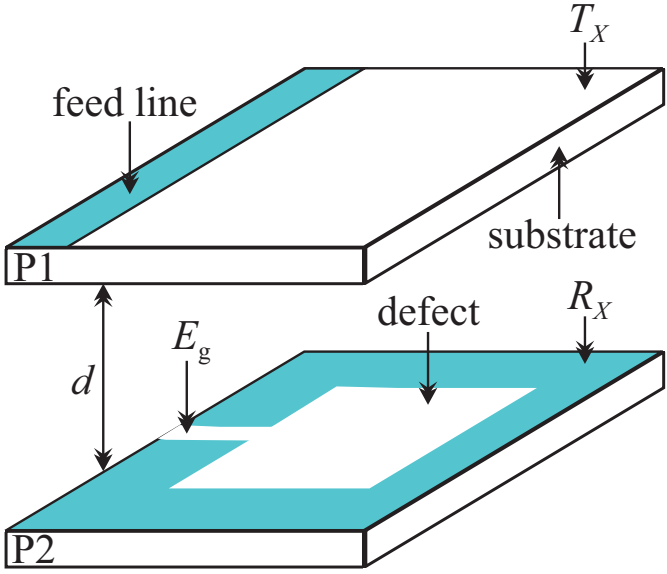}
	\caption{Coupling example of two DGS-based resonators. P1 and P2 stand for the input and output ports, respectively. Note that the defect type can be replaced by any shape. }
	\label{fig: coupling}
\end{figure}

\begin{figure}[!t]
	\centering
	\includegraphics[width=1\linewidth]{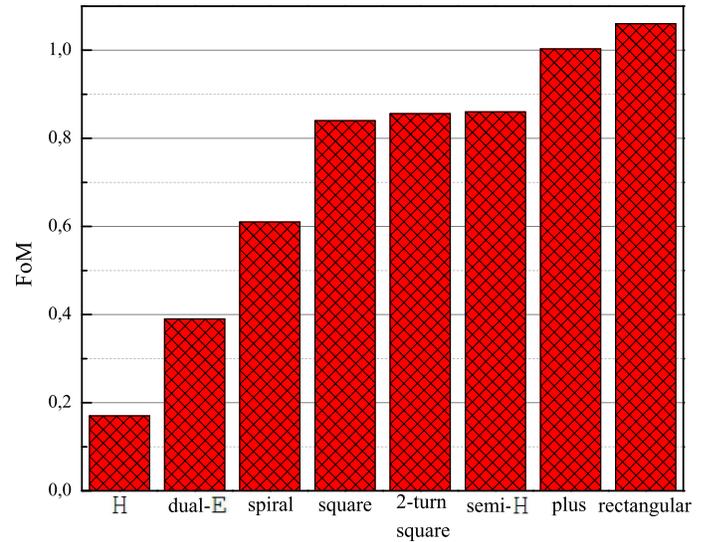}
	\caption{The achieved FoM values of the single-band WPT systems employing different DGS shapes.}
	\label{fig: FoM}
\end{figure}

The realization of the WPT systems requires the utilization of two coupled (see Figure \ref{fig: coupling}) antennas, acting roles of a transmitter and a receiver, both resonating at the same frequency. There has been an upward spike in the use of the DGS-based resonators in the WPT systems as it leads to circuit size reduction~\cite{Hekal-5,Chalise-1}. The first report of DGS-based WPT was developed around \verb|H|-shaped resonators~\cite{Hekal-2}. Subsequently, various analyses and design approaches to enhance this work both in terms of performance as well as ease of design have been reported \cite{Kupreyev-1,Verma-2}. In particular, the \textit{J-}inverter theory seems very promising as it readily aids in the analysis and extraction of the WPT equivalent circuit element values \cite{Verma-1,Tahar-1}. Some reported works showed the impact of misalignment between $T_{X}$ and $R_{X}$ by presenting numerous studies~\cite{Barakat-1,Hekal-1}. Furthermore, the literature is also replete with recent papers which broadened the WPT schemes to both wireless power and information transfer by proposing multi-band WPT systems~\cite{Barakat-3,My-2}. Then, there have been efforts to benchmark the WPT systems by proposing standard metrics such as the figure-of-merit (FoM). The FoM essentially includes all the key parameters of WPTs, namely, power transfer distance ($d$), circuit size, and efficiency ($\eta$) of the delivered power to $R_{X}$. In addition, the further advancements in the variously shaped single-band DGS WPTs are compared in terms of FoM (see Figure \ref{fig: FoM}) and the other parameters are listed in Table \ref{tab: WPT comparison}.    

As mentioned earlier, $\eta$ is one of the critical parameters of WPTs that heavily depends on the resonator's inductance ($L$), \textit{i.e.}, a higher value of $L$ results in efficiency enhancement. This phenomenon was examined through different shapes of DGSs, such as \verb|H|, semi-\verb|H|, and two-turn square DGS~\cite{Hekal-4,Hekal-1}. It was shown that semi-\verb|H| shape has greater inductance as compared to \verb|H| shape because of possessing only one current path and, therefore, the WPT systems employing this resonator exhibit higher efficiency. Moreover, two-turn square DGS also has higher inductance in contrast to \verb|H| shape because of the extended current loop and, therefore, its use also enhances the WPTs efficiency. Moving forward, the positive influence of $Q$ on a system efficiency was demonstrated by exploiting the dual-\verb|E| shape DGS~\cite{Verma-1}. It was also shown, that the efficiency of WPTs is severely impacted by the propagation medium and, therefore, it should be taken into account for any practical situation~\cite{Chalise-1}. In addition, the efficiency is greatly impacted by misaligned $T_{X}$ and $R_{X}$ resonators and it was investigated using a semi-elliptic DGS~\cite{Barakat-1}. The efficiency also gets enhanced substantially in multi-band WPT systems as well which is apparent from the developed dual-band WPT system based on cascades of a circular DGS~\cite{Tahar-1}.

The power transmission range is another very important aspect of any WPT system and, therefore, a number of reports, based around a variety of DGS shapes, tried to improve $d$. It has been established that $d$ is directly proportional to the value of a coupling coefficient ($k$), \textit{i.e.} higher $k$ leads to a higher transmission range and vice-versa~\cite{Tahar-2}. It was demonstrated through the use of rectangular DGS resonators which also possess a higher $Q$ value. In addition, other reported WPT systems, that focus on a range enhancement, considered the plus and hexagonal shapes of DGSs~\cite{My-1,Kupreyev-1}. 

\begin{table}[]
	\footnotesize
	\centering
	\caption{Performance comparison of the DGS-based WPT systems ($^{\star}-$single-band, $^{\circ}-$dual-band, and $^{\dagger}-$tri-band).}
	\label{tab: WPT comparison}
	\begin{tabular}{|c|c|c|c|c|c|}
		\hline
		Ref. & \begin{tabular}[c]{@{}c@{}}$\eta$ \\ (\%)\end{tabular} & \begin{tabular}[c]{@{}c@{}}$d$\\ mm\end{tabular} & \begin{tabular}[c]{@{}c@{}}size \\ mm$ \times$mm\end{tabular} & FoM \\ \hline
		\cite{Hekal-2}$^{\star}$ & $85$ & $5$ & $25\times25$ & $0.17$ \\ \hline
		\cite{Hekal-4}$^{\star}$ & $68.5$ & $50$ & $40\times40$ & $0.856$ \\ \hline
		\cite{Hekal-1}$^{\star}$ & $73$ & $25$ & $21\times21$ & $0.86$ \\ \hline
		\cite{Verma-1}$^{\star}$ & $71$ & $11$ & $20\times20$ & $0.39$ \\ \hline
		\cite{Chalise-1}$^{\star}$ & $61$ & $50$ & $50\times50$ & $0.61$ \\ \hline
		\cite{Hekal-5}$^{\star}$ & $63$ & $40$ & $30\times30$ & $0.84$ \\ \hline
		\cite{Tahar-2}$^{\star}$ & $64.6$ & $44$ & $35.8\times20$ & $1.06$ \\ \hline
		\cite{My-1}$^{\star}$ & $86$ & $35$ & $30\times30$ & $1.003$ \\ \hline
		\cite{Tahar-1}$^{\circ}$ & $71/72$ & $16$ & $30\times15$ & $0.757/0.772$ \\ \hline
		\cite{My-2}$^{\circ}$ & $71/81$ & $15$ & $18\times18$ & $0.84/0.96$ \\ \hline
		\cite{Verma-2}$^{\circ}$ & $60/67$ & $3.5$ & $15\times10$ & $0.17/0.19$ \\ \hline
		\cite{Kupreyev-1}$^{\circ}$ & $58/74$ & $17.5$ & $34\times20$ & $0.55/0.69$ \\ \hline
		\cite{Barakat-4}$^{\circ}$ & $70/69$ & $40$ & $50\times50$ & $0.56/0.55$ \\ \hline
		\cite{Sharaf-2}$^{\circ}$ & $80/73$ & $17$ & $20\times20$ & $0.68/0.62$ \\ \hline
		\cite{Barakat-3}$^{\dagger}$ & $68/60/65$ & $30$ & $50\times50$ & $0.4/0.36/0.39$ \\ \hline
	\end{tabular}
\end{table}

Miniaturization of WPTs is one of the most important aspects of the development of the DGS-based WPT systems for biomedical and other low-power applications. Recently reported articles tried to address this requirement. For instance, an asymmetric resonator-based WPT system, where $R_{X}$ is smaller than $T_{X}$, is one solution to reduce the overall system size~\cite{Hekal-5}. Other exciting designs for size miniaturization include the ultra-compact \verb|U| and double-rectangular DGS shape dual-band WPT systems working at practical industrial, scientific, and medical (ISM) bands~\cite{Verma-2,My-2}. Recently, the concept of multi-mode DGS resonators has been incorporated for achieving compact-sized dual- and tri-band WPT systems~\cite{Barakat-4,Barakat-3}. In addition, it has been shown that the size miniaturization of dual-band WPT can also be achieved by adding a lumped capacitor to resonators~\cite{Sharaf-2}.

\section{DGS-based WPT Design Considerations}
\label{sec: WPT Challenges}
Apart from the main parameters of WPTs, namely, efficiency, resonator compactness, and power transmission distance, the design of DGS-based WPTs faces inherent challenges and, therefore, requires appropriate design techniques for optimal performance. The key considerations during the design trade-offs are within the realm of the resonator design, equivalent circuit development, impedance matching, and high $k$ and $Q$ realization. The overall idea is to develop a precise methodology to design and realize the WPT systems possessing extremely high FoM with respect to given applications.

\subsection{DGS Technique}
\label{subsec: Resonator_DGS}
The photonic bandgap structure (PBG) or DGS techniques are capable of achieving stop-band or band-gap effects and, hence, are extremely useful in the design and development of microwave and millimeter-wave filtering circuits \cite{Lim,Huang-1}. The rejection of the desired frequencies in PBG is regulated by the number of lattices, lattice spacing, and relative volume fraction, whereas it has been well established that in the DGS technique this is primarily dependent on the physical dimensions of the etched shape in the ground plane \cite{Kim-2,Ahn-1}, as the DGS concept lies in altering the current distribution of the ground plane by introducing any type of defect on it. As a consequence, the capacitance and the inductance of the transmission line change. In general, DGS possesses numerous advantages such as the ease of modeling, simplicity in constructing the equivalent circuit, fabrication, \textit{etc.}, hence is preferred over PBG in the design of filtering structures such as band-stop filters (BSFs) \cite{Huang-1}, low-pass filters \cite{Mandal-1}, and band-pass filters \cite{Abdel-Rahman-1}. Analysis of any DGS shape is extremely difficult and, therefore, it is standard practice to make use of advanced electromagnetic (EM) simulators to solve the DGS related challenges \cite{Liu-1}. Moreover, the parameters of the DGS-based resonators can be extracted by using a simple $LC$  tank circuit, but it cannot provide information about the physical dimensions of the introduced DGS shape and, hence, this approach cannot be considered unique for any DGS shape. For the parameter extraction of a particular DGS shape, the development of a quasi-static equivalent circuit is required \cite{Karmakar-1}. 

Furthermore, the available literature includes diverse analyses of the DGS technique from multiple aspects. For instance, the authors in \cite{Woo} improved the performance of the circuit by introducing \verb|U| and \verb|V| shaped DGSs compared to the earlier reported dumbbell and spiral shapes. The resonator miniaturization was achieved by utilizing bending feasibility of the microstrip in \cite{Liu-4}, while \cite{Wang} reported a $94\%$, $57\%$, and $88\%$ size reduction compared to the conventional dumbbell, \verb|H|, and modified dumbbell shape-based BSFs, respectively, by embedding varactors to the circuit to enable tunable resonators. Asymmetric types of DGS were first explored in \cite{Kumar}, where the authors concluded that this type of DGS shapes outperforms the symmetric DGS shapes in terms of the resonator size reduction. Another tunable band-stop resonator, possessing a small circuit size, employs \verb|E| shape DGS which has a high $Q$ and a sharp transition knee \cite{Huang-1}.        
 
\begin{figure}[!t]
	\centering
	\subfloat[circle]{
		\label{subfig: circle}
		\includegraphics[width=0.3\columnwidth]{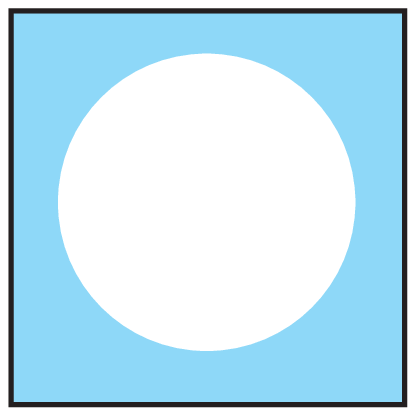}}~
	\subfloat[ellipse]{
		\label{subfig: ellipse}
		\includegraphics[width=0.3\columnwidth]{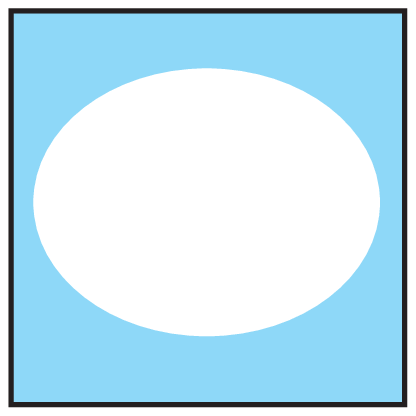}}~
	\subfloat[\texttt{H}]{
		\label{subfig: triangle}
		\includegraphics[width=0.3\columnwidth]{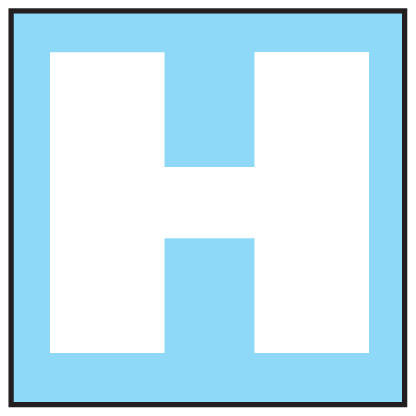}}\\
	\subfloat[plus]{
		\label{subfig: plus}
		\includegraphics[width=0.3\columnwidth]{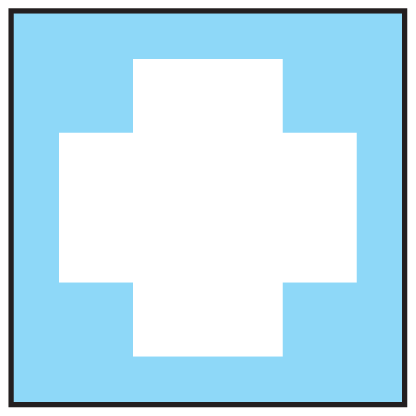}}~
	\subfloat[square]{
		\label{subfig: square}
		\includegraphics[width=0.3\columnwidth]{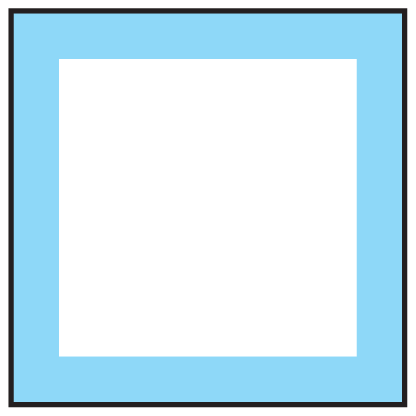}}~
	\subfloat[\texttt{T}]{
		\label{subfig: T}
		\includegraphics[width=0.3\columnwidth]{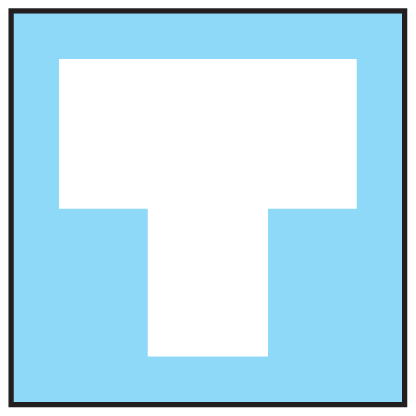}}~
	\caption{The utilized different DGS shapes.}
	\label{fig: DGS shapes}
\end{figure}

\begin{figure}[t!]
	\centering
	\includegraphics[width=1\linewidth]{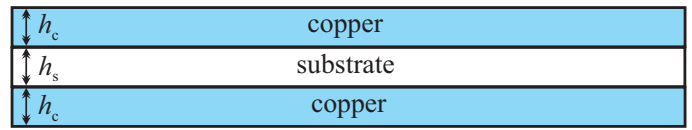}
	\caption{The structure of the DGS-based resonators.}
	\label{fig: substrate structure}
\end{figure}

\subsection{DGS shape analysis}
\label{subsec: Different DGS shape analysis}
\begin{figure}[t!]
	\centering
	\includegraphics[width=1\linewidth]{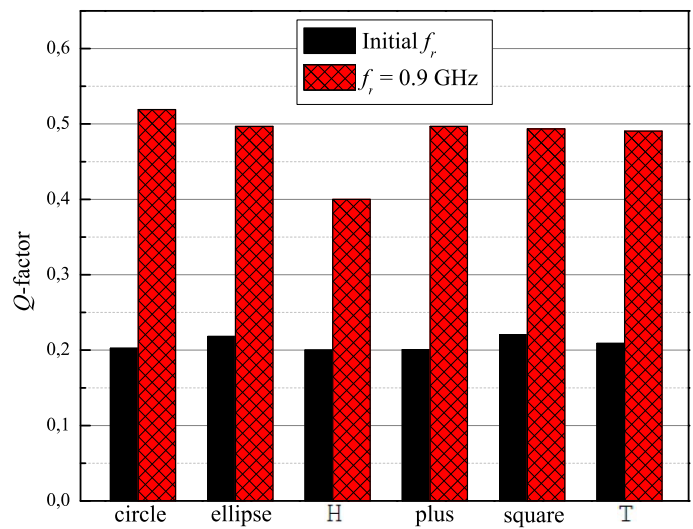}
	\caption{$Q$ factor of resonators based on various considered DGS shapes.}
	\label{fig: Q-factor}
\end{figure}

One approach of achieving a highly-efficient WPT system is to improve the $Q$ value of a resonator \cite{Tahar-2,Imura-1}. Therefore, this section provides analyses of various-shaped DGS-based resonators in terms of the $Q$ value. As mentioned above, different shapes of DGS in the ground plane disturb the current distribution in varying fashion and it leads to numerous band-stop characteristics. To visualize this aspect, a brief survey was carried out and some of the common DGS shapes, shown in Figure~\ref{fig: DGS shapes}, have been selected to develop BSF and analyze them in terms of $Q$. A typical resonator structure used in WPT systems is given in Figure \ref{fig: substrate structure}. For the sake of precise comparison, all the resonators have been developed using the same substrate and have the same board area of $20\times20$ mm$^{2}$. For completeness of these simulations, the substrate RO$4350$B was selected with parameters of copper thickness $(h_{\text{c}})= 35~{\mu}$m, substrate thickness $(h_{\text{s}})=1.524$ mm, and a substrate permeability $(\varepsilon)=3.66$. The ML (\textit{i.e.}, feed line), located just below the excitation gap ($E_{\text{g}}$), connects the input and output ports of resonators (as shown in Figure \ref{fig: coupling}). It is important to mention that although the resonators possess different DGS shapes in the ground plane they are simulated under the same conditions, \textit{i.e.}, a ratio of the copper area to the non-copper one was set to $1:1$ for all cases. It can be inferred from the obtained results in Figure \ref{fig: Q-factor} that the \verb|H| shaped resonator has the smallest $Q$ value compared to the resonators of other shapes. However, it is apparent that there is no huge difference in the $Q$ value among the chosen shapes. Subsequently, it is reasonable to conclude that the shape is not that important in achieving high $Q$.

\subsection{Equivalent Circuit}
\label{subsec: Resonator_Resonator Equivalent Circuit}
It has been demonstrated in \cite{Tahar-1,Verma-1} that the validation of the DGS-based WPT systems is performed by investigating their corresponding equivalent circuits. The equivalent circuit of any DGS-based WPT system is preceded by the development of the equivalent circuit of the employed resonator (see Figure \ref{subfig: resonator eq circ}). The respective parameters of this equivalent circuit require the usage of analytical expressions \cite[Eqs. (1) and (2)]{My-2}. In general, any WPT system is developed by coupling two resonators and, as a consequence, a WPT system can be represented by their equivalent circuits coupling with each other as depicted in Figure \ref{subfig: WPT eq circ}. Here, open stub ($C_{\text{st}}$) is added to fulfill the impedance matching and its detailed calculation can be found in \cite{Hekal-1}. In this instance, it should be noted that there are two ways to design a multi-band resonator, by either cascading single band-resonators~\cite{Tahar-1,My-1} or by utilizing multi-mode techniques~\cite{Barakat-4,Barakat-3}.  

\subsection{Methodology}
\label{subsec: Methodology}
\begin{figure}[t!]
	\centering
	\includegraphics[width=1\linewidth]{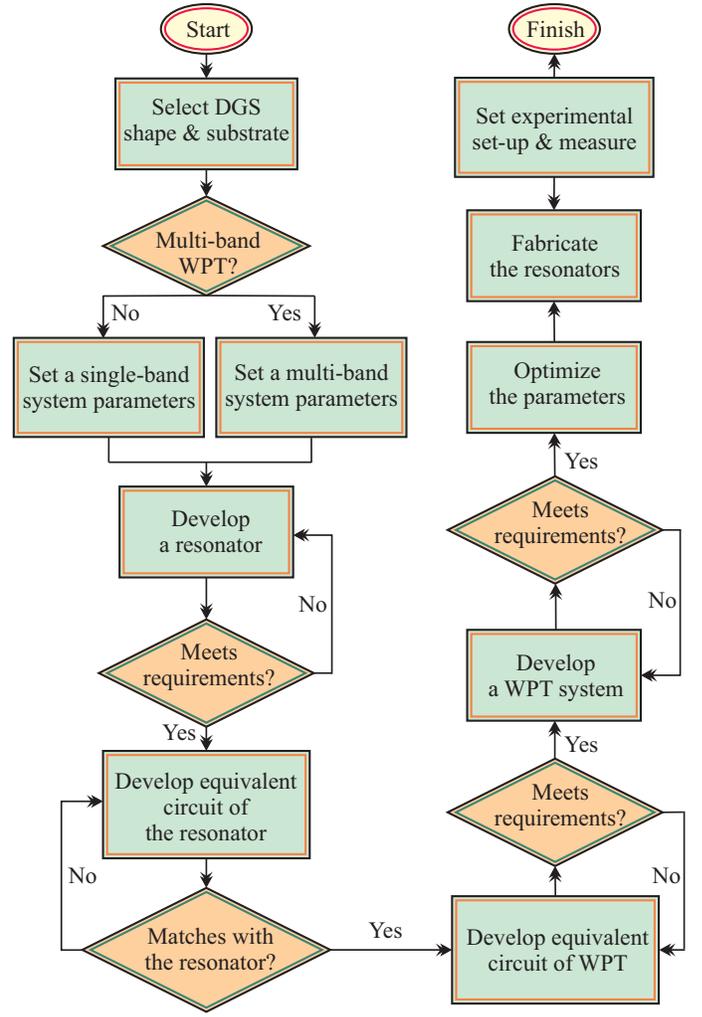}
	\caption{The systematic methodology of the DGS-based WPT system development.}
	\label{fig: flow chart}
\end{figure}

The successful realization of any practical WPT system requires the simulation and experimental demonstration. Over the years, several techniques to design DGS-based WPT systems have been reported in \cite{Sharaf-2,Hekal-4,Tahar-1}. These techniques, although, differ in implementation strategies but can still be described by a standard design procedure as structured by a flow chart depicted in Figure \ref{fig: flow chart}. For the sake of completeness, the design steps are elaborated below: 

\begin{itemize}
	\item At the outset, define a DGS shape and select a substrate for the intended design. For example, any shape from Figure \ref{fig: DGS shapes} (but not limited) and the substrate RO$4350$B. 
	
	\item Check the requirement of whether the design is multi-band or single-band. For instance, single-band is required only for power transfer whereas a multi-band is envisaged for power as well as information transfer. However, care should be taken so that the chosen frequencies conform to the approved standards, \textit{e.g.}, ISM. 
	\item Then, specify the key parameters of the WPT system, namely, operating frequency, resonator area, and power transfer distance according to the requirements of the planned application.
	
	\item Make use of the chosen DGS shape and the specified WPT parameters to develop a resonator. Calculate the value of an external capacitor, using \cite[Eq. (2)]{My-2}, to achieve the resonance at a chosen frequency. For multi-band resonance, either cascade multiple resonators \cite{Tahar-1,Kupreyev-1,My-2} or follow the multi-mode techniques\cite{Barakat-4,Barakat-3}. It is important to check if the resonance requirements are satisfied. 
	
	\item Subsequently, develop an equivalent circuit of the resonator to validate the obtained EM simulation results.  
	
	\item In turn, to construct the WPT system's equivalent circuit, first, there is a need to couple two identical resonators and separate them by distance $d$. To fulfill this necessity, carefully place two resonators in a way that no misalignment occurs between them and ensure that they remain under the perfect coupling condition. Then, use \cite[Eqs. (3) and (4)]{Verma-2} to extract the values of $M$ and $k$, respectively. Finally, develop an equivalent circuit of the WPT system using those defined values.     
	
	\begin{figure}[t!]
		\centering
		\subfloat[resonator]{
			\label{subfig: resonator eq circ}
			\includegraphics[width=1\columnwidth]{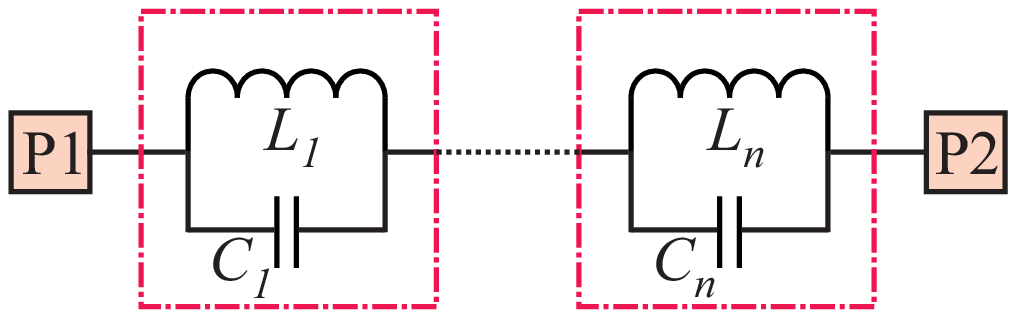}}\\
		\subfloat[WPT system]{
			\label{subfig: WPT eq circ}
			\includegraphics[width=1\columnwidth]{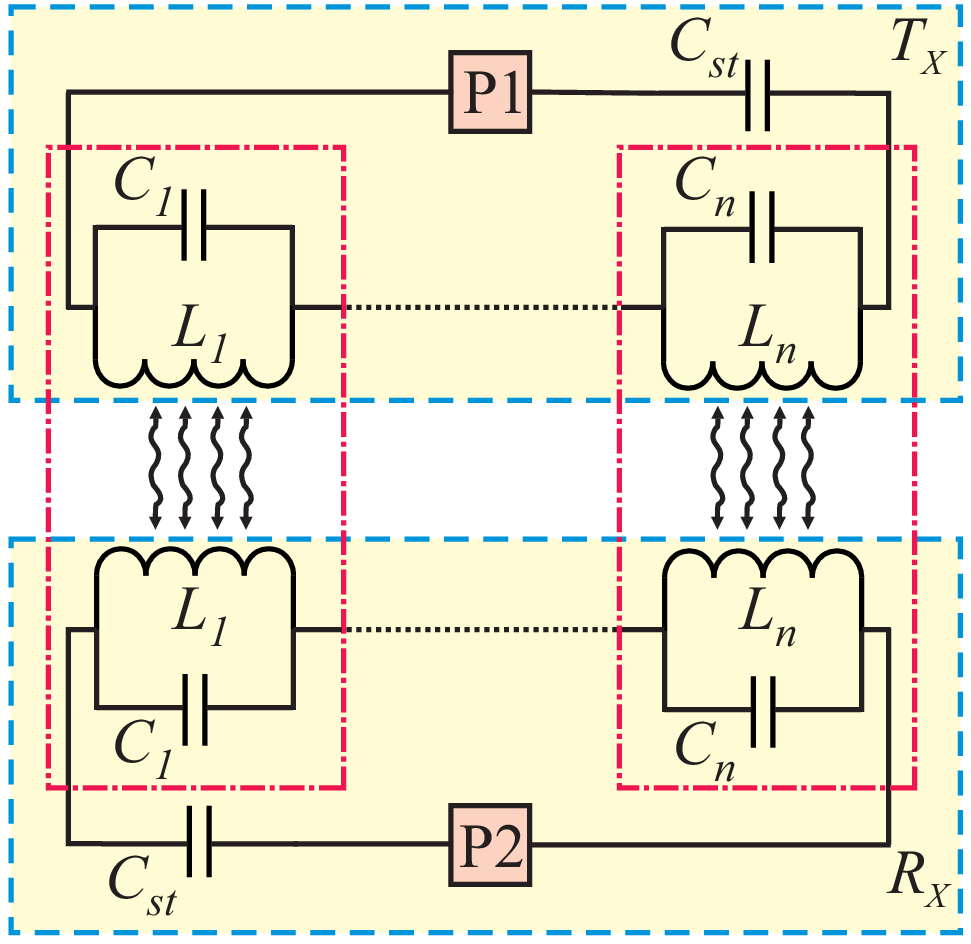}}
		\caption{Equivalent circuits.}
		\label{fig: Equiv circuits}
	\end{figure}
	
	\item Perform impedance matching of the circuit, \textit{i.e.}, define the value of $C_{\text{st}}$, once the equivalent circuit of WPT is developed. Next, convert $C_{\text{st}}$ into open-stub ML \cite[Eq. (4)]{Hekal-1} and then develop the WPT system.  
	
	\item Optimization of critical parameters of WPT is often necessitated to achieve the best performance and often to trade-off among performance metrics. This completes the simulation phase. 
	
	\item Fabricate two resonators to initiate the experimental demonstration and solder the required lumped components of the resonators (the connectors and capacitors).
	
	\item Perform measurements using network analyzers. Store data and use (\ref{eq: efficiency}) \cite{My-1} and (\ref{eq: FoM}) \cite{Sharaf-2} to calculate the well-accepted performance metrics of the WPT systems. However, prepare the experimental setup carefully for the precise results. A good agreement between the experimental and simulation results validates the design process and the realized DGS-based WPT system. 
	
	\begin{equation}
	\centering
	\label{eq: efficiency}
	\eta = \frac{\mid S_{21}\mid^2}{1-\mid S_{11}\mid^2}, 
	\end{equation}
	
	\begin{equation}
	\centering
	\label{eq: FoM}
	\text{FoM}=\eta\times\frac{d}{\sqrt{\text{average size}}}.
	\end{equation}
	
\end{itemize}

\begin{figure*}[t!]
	\centering
	\subfloat[resonator]{
		\label{resonator}
		\includegraphics[width=1\columnwidth]{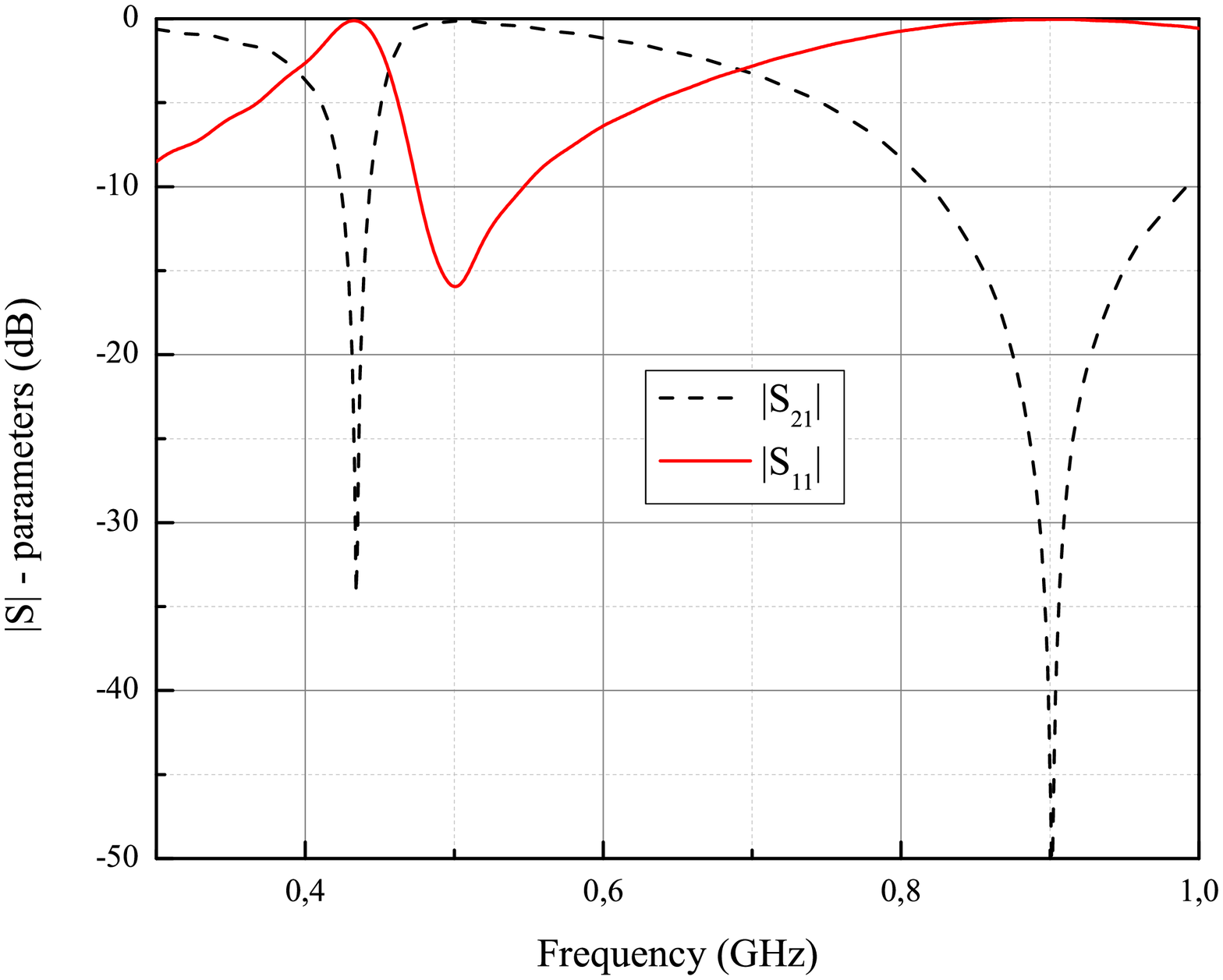}}
	\subfloat[WPT system]{
		\label{WPT syst}
		\includegraphics[width=1\columnwidth]{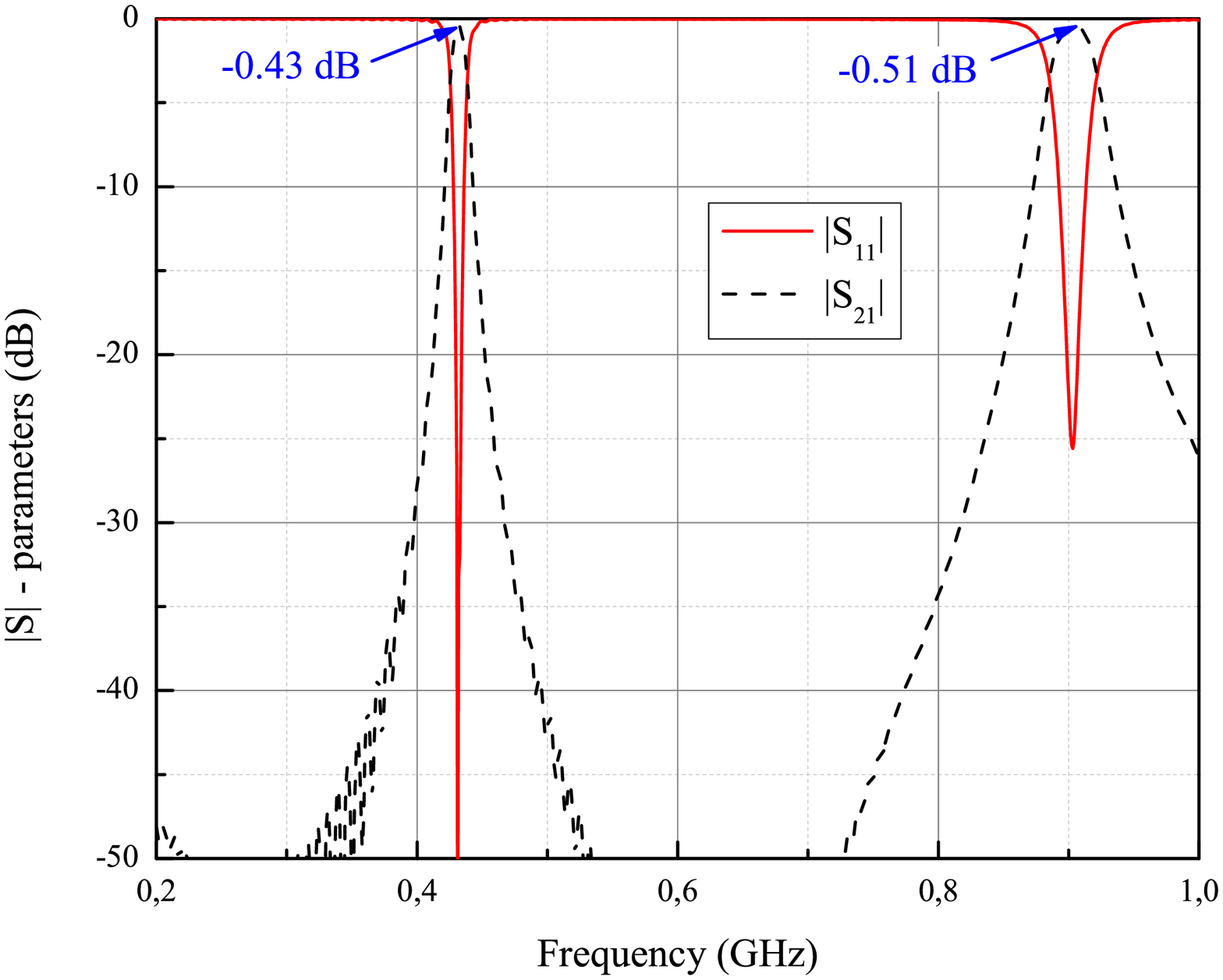}}
	\caption{The obtained $\left| \text{S} \right|$-parameters from EM simulations \cite{My-2}.}
	\label{Sim results}
\end{figure*}

\section{Realized DGS-based WPT System Examples}
\label{sec: Realized WPTs}
The aim of this section is to introduce the readers with some recently reported exciting DGS-based WPT designs which could bring a paradigm shift in this domain. At first, an example of the design of a dual-band WPT system operating at $0.433$ GHz and $0.9$ GHz is presented to demonstrate the effectiveness of the existing methodology. It can be seen in Figure \ref{fig: coupling} that, in general, two identical resonators with a certain separation distance between them need to be coupled to develop a dual-band WPT system. Particularly, it is critical to define the operating distance carefully so that the two resonators are under a perfect coupling condition. For example, the simulation of the coupled dual-band resonators defines the WPT system results in $\left| \text{S}\right|$-parameters depicted in  Figures \ref{resonator} and \ref{WPT syst}. It can be deduced from these plots that there are two distinct resonant frequencies and they can be considered as the operating frequencies of the WPT system. These results clearly show that the DGS-based resonators are useful in realizing WPT systems. It is important to mention that these frequencies can be easily regulated by placing an external capacitor to the utilized resonators.

The next step, as mentioned in Section \ref{subsec: Resonator_Resonator Equivalent Circuit}, entails the development of the equivalent circuit model of the resonators and corresponding WPT system for validation of the achieved results. The two-step validation gives confidence in the designed WPT system before eventual fabrication. It is anticipated that the measured values of $\left| \text{S}\right|$-parameters will follow the simulation results. In case of any anomaly, slight adjustments in the design stage need to be carried out along with the care in the measurement. From the achieved plots, in this case, equations (\ref{eq: efficiency}) and (\ref{eq: FoM}) give the efficiencies of $90\%$ and $88\%$ at the two chosen frequencies while the corresponding FoMs are $1.06$ and $1.03$.  

\begin{figure*}[t!]
	\centering
	\subfloat[\texttt{H}-shaped single-band WPT \cite{Hekal-1}]{
		\label{subfig: H-shape DGS realization}
		\includegraphics[width=0.65\columnwidth]{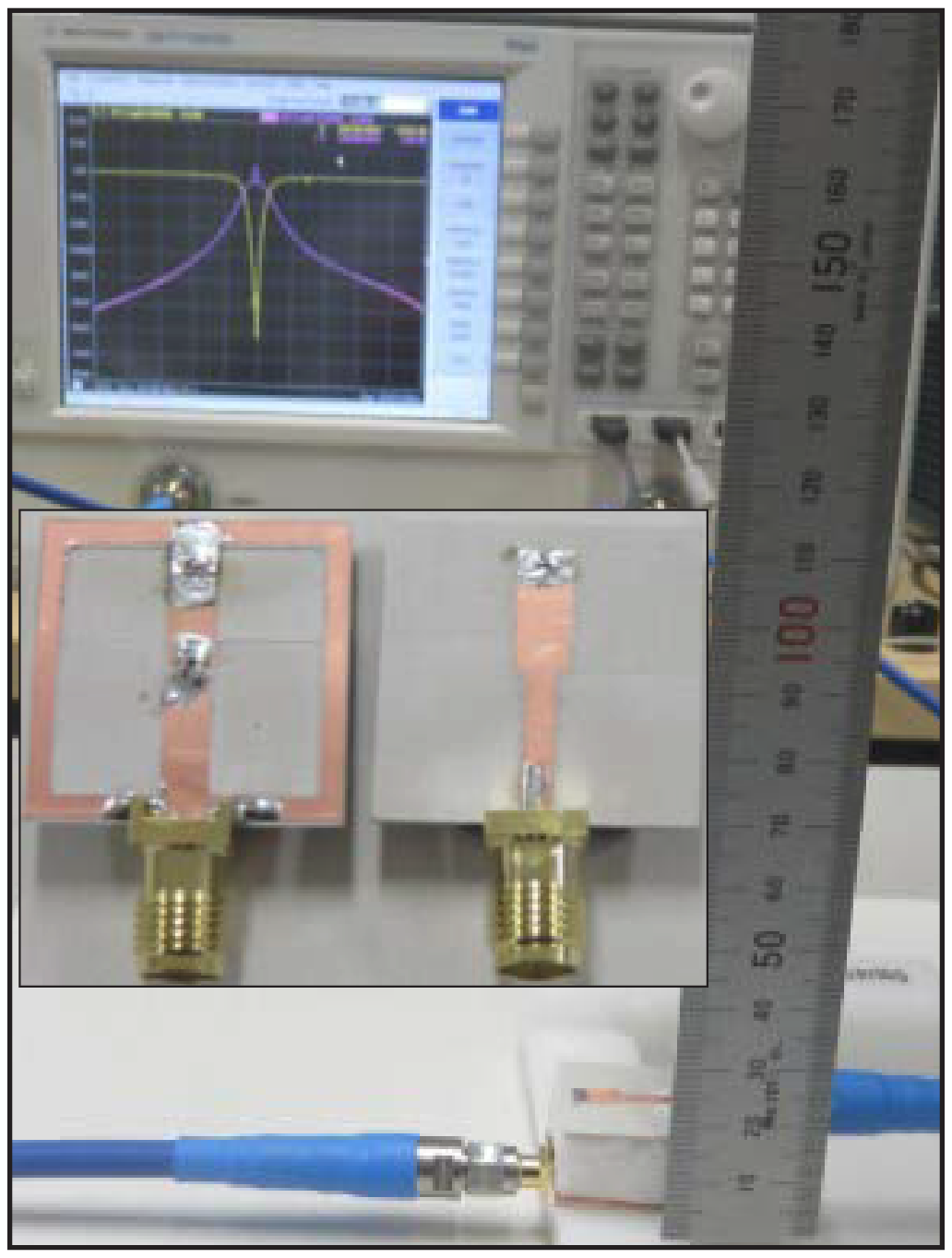}}~
	\subfloat[double-square shaped dual-band WPT \cite{My-2}]{
		\label{subfig: dual realization}
		\includegraphics[width=0.61\columnwidth]{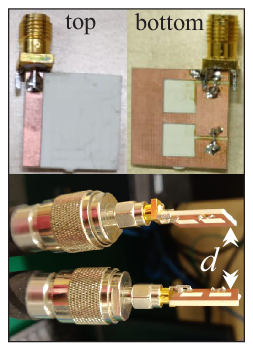}}~
	\subfloat[multi-mode based tri-band WPT \cite{Barakat-3}]{
		\label{subfig: tri_realization}
		\includegraphics[width=0.603\columnwidth]{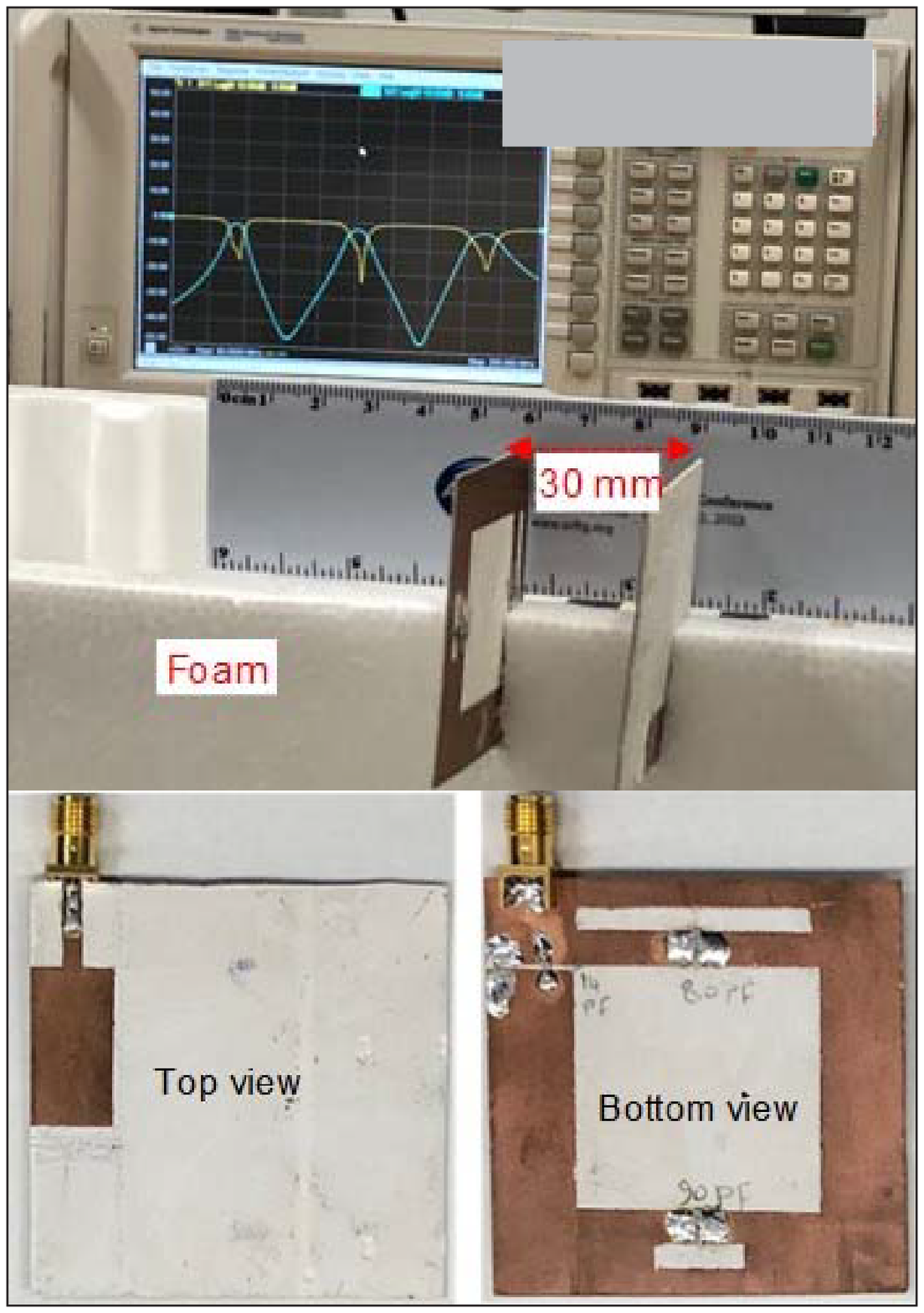}}
	\caption{Successfully realized DGS-based WPT systems: (a) single-band WPT; (b) dual-band WPT; (c) tri-band WPT.} 
	\label{fig: Realized WPTs}
\end{figure*}

Over the years there have been numerous successful realizations of the DGS-based WPT systems. For example, one such design reported in \cite{Hekal-1} aroused a keen interest in the DGS-based WPT system. This design, depicted in Figure \ref{subfig: H-shape DGS realization}, was considered very advanced at the time of its reporting. It provided an efficiency of $73\%$ at the designed frequency of $0.3$ GHz. Similarly, a recent design, depicted in Figure \ref{subfig: dual realization}, delves into the dual-band WPT system for practical ISM bands and has the potential for its usage in real-life scenarios \cite{My-2}. Another multi-band design, given in Figure \ref{subfig: tri_realization}, is the first time a tri-band WPT system \cite{Barakat-3} has been presented and it definitely opens the door for real-time applications of such systems in simultaneous power and information transfer. 

\section{Ongoing Research Investigations}
\label{Sec: Future Works}
The recent reports in the field of DGS-based WPT systems address the issues related to the main challenges such as power transfer efficiency, distance, circuit size, high $Q$ realization, \textit{etc}. For instance, most of the papers often reported the analysis and design process and overlooked the issues encountered in the overall WPT system design. In particular, some DGS-based designs provided peripheral analysis using their quasi-static equivalent circuits~\cite{Hekal-5,Hekal-1}, but most of the designs did not employ this approach and, hence, it may not be possible to understand the anomaly in their respective EM and circuit simulation results. In particular, such an inference can be drawn from most of the recent designs~\cite{Verma-1,Hekal-4} due to the absence of any unified approach adopted for the development of the WPT system's equivalent circuit. This fact opens up an opportunity for advanced research in developing a standardized approach for the WPT equivalent circuit. This has the potential to assist performance validation prior to the fabrication stage and, thus, can aid in the successful first-pass WPT design. As mentioned previously, the operating distance and power transfer efficiency in such systems can be accomplished by high $Q$ resonators, but none of the realized WPT systems reported so far achieved more than $85\%$ efficiency \cite{Hekal-4,Barakat-4}. This again throws an open challenge to designers to tread this path of high $Q$ resonators design with target applications in the WPT systems. Furthermore, the misalignment between $T_{X}$ and $R_{X}$ has a severely detrimental effect on the efficiency of WPT as reported in~\cite{Hekal-2,Barakat-1}. This aspect needs to be investigated thoroughly to identify and mitigate the impact of misalignment on the WPT performance. Another factor, known as a frequency splitting, also degrades the efficiency and it can be seen in some reported papers \cite{Tahar-2,Sharaf-2}. Accordingly, this phenomenon can be considered as another interesting research path. 

Several reported WPTs were designed to operate at frequencies that do not comply with standards such as WBAN or ISM  \cite{Barakat-4,Barakat-3}. Furthermore, there are not many reports which demonstrate the performance of WPTs in real tissue environments and, therefore, it is extremely difficult to make any judgment on their field deployments \cite{Chalise-1}. Thus, this is another open research direction for the eventual incorporation of WPTs in the biomedical field. The review results reveal that there is a need for new performance metrics for such systems that can also be applied to any WPT.

Keeping the above in mind, there is a requirement of innovative designs covering all aspects of the design process as well as the performance metrics despite the fact that the DGS-based WPT systems have gone through considerable advancements in this decade. In this context, the potential research concerns of the DGS-based WPT systems can be categorized into four main groups as shown in Figure \ref{fig: research directions}. The following sub-sections elaborate on these concerns, the associated challenges, and the possible directions based on the available knowledge. 

\begin{enumerate}
	\item Performance Enhancement 
	\begin{enumerate}
		\item The WPT system's efficiency actually regulates the amount of power (received by $R_{X}$) to satisfy the needs of a wide range of applications. Here, the aim must be set to achieve an efficiency of at least $90\%$ so that the majority of the transmitted power is received by the load side.
		
		\item Another critical aspect that requires considerable attention is the compactness of WPT systems considering that it will facilitate several applications such as charging of IMD batteries. 
		
		\item The power transmission range is also very important for WPT systems. For example, consumer electronics may often benefit from the DGS-based WPT systems possessing a relatively larger power transmission range. In general, the large-sized WPT systems exhibit larger operating distance and, therefore, there exists a trade-off to achieve the optimal size and transmission range. Therefore, efforts are being put in this direction to make these trade-offs application-centric. 
		
		\begin{figure*}[t!]
			\centering
			\includegraphics[width=1\linewidth]{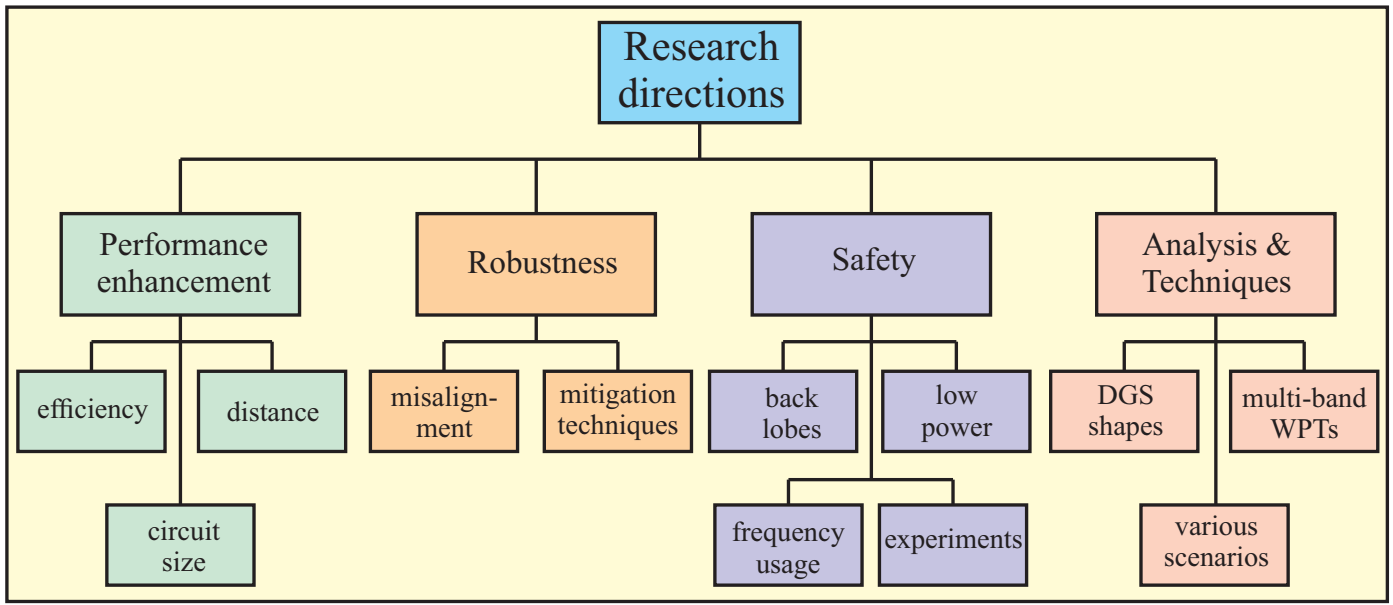}	\caption{The research directions in the domain of the DGS-based WPT systems.}
			\label{fig: research directions}
		\end{figure*}    
		
	\end{enumerate}
	\item Robustness
	\begin{enumerate}
		\item A system that works repetitively without too much hassle is the need of the time. Therefore, the robustness of the DGS-based WPT systems is another aspect that must be the focus in the near future. Specifically, different types of misalignment between $T_{X}$ and $R_{X}$ such as angular, horizontal, and vertical can degrade the WPT system performance considerably. New methodologies are required to account for these issues and provide robustness to the WPT systems.  
		
		\item It is often reported that a disagreement occurs between the simulation and experimental results in such systems. Thus, there is a pressing need to develop a new methodology, such as integrated hardware in the loop type solution, that can mitigate this issue.
		
	\end{enumerate}
	\item Safety
	\begin{enumerate}
		\item One of the main applications of the near-field WPT systems relates to the broad biomedical domain. In this context, the safety aspect is paramount as it concerns the use and handles by a human for humans. The idea here should be to avoid any adverse impact on human health. It is important to mention that the existing solutions have not considered this issue but it is pertinent to investigate this further so as to develop techniques for reducing the back-lobe radiations of the resonators.  
		
		\item The high-power WPT systems may not be appropriate for numerous applications including IMDs, for safety reasons, and, in such cases, the low-power near-field WPT systems are mostly employed. Hence, the innovative WPT systems with extremely high efficiency are the need of the time.
		
		\item Most of the existing works report the designs operating at arbitrary frequencies but these may not be very appropriate for the biomedical domain considering that it concerns human health. Accordingly, future research should consider only the Federal Communications Commission (FCC) approved reserved frequency bands as it helps to standardize the concerns associated with the WPT systems employed in the biomedical field.   
		
		\item It is worthwhile mentioning that the available literature still lacks the reports considering the performance of the DGS-based WPT system in the real tissue environment.  Consequently, more experiments and explorations of power transfer through biological tissues are needed. 
	\end{enumerate}
	
	\item Design and Analysis Techniques 
	\begin{enumerate}
		\item So far, the existing solutions have been relied on introducing different-shaped defects on the ground plane to achieve the desired resonant frequency and obtain high $Q$. It has been demonstrated in Section \ref{subsec: Different DGS shape analysis} through the case studies that different shapes possess almost the same $Q$. Besides, the varying resonant frequency can be tuned using an external capacitor. Accordingly, this approach can be definitely considered as a sub-optimal design process. Therefore, to further advance the design approaches, the focus should not be on various shapes but on the development of some techniques based on quasi-static equivalent circuits. Furthermore, there is a need to support designs through closed-form equations relating to $M$ and $k$ for achieving scalability and repeatability. This will eventually assist in the overall planning of the desired WPT system prior to starting the eventual design.  
		\item Besides, new techniques to design the compact multi-band WPT systems with excellent performance have the potential to bring a paradigm shift in a broad range of the internet of things (IoT) devices and therefore should be prioritized.
		
		\item Other design scenarios that have the potential to enhance the power transfer range and improve other design metrics that should also be considered. For example, a multi-hop scenario, \textit{i.e.}, placing an intermediate resonator between $T_{X}$ and $R_{X}$, seems a good direction to pursue advanced research as it may provide a huge improvement in terms of $d$.
	\end{enumerate}
\end{enumerate}

\section{Conclusions}
\label{Conclusion}
This paper provides a comprehensive resource about the challenges and recent advancements in the domain of the DGS-based WPT systems. Analysis of the different examined  DGS shapes shows that the average $Q$ value can be obtained by utilizing any of them. Although this field has been advancing in the last several years this area still needs some research considerations, \textit{e.g.}, human safety aspects, new design techniques, different analyses, \textit{etc}. In addition, the provided review reveals that the WPT systems have been broadened to the multi-band structures enabling the system to transfer power and data simultaneously.

\balance


\begin{thebibliography}{99}
	
	\bibitem{Ng}
	W. M. Ng, C. Zhang, D. Lin and S. Y. Ron Hui, \textquotedblleft Two-and Three-Dimensional Omnidirectional Wireless Power Transfer,\textquotedblright~\textit{IEEE Transactions on Power Electronics}, vol. 29, no. 9, pp. 4470-4474, September 2014.
	
	\bibitem{Kung-2}
	M. Kung and K. Lin, \textquotedblleft Dual-Band Coil Module With Repeaters for Diverse Wireless Power Transfer Applications,\textquotedblright~\textit{IEEE Transactions on Microwave Theory and Techniques}, vol. 66, no. 1, pp. 332-345, January 2018.
	
	\bibitem{Jonah-1}
	O. Jonah and S. V. Georgakopoulos, \textquotedblleft Wireless Power Transfer in Concrete via Strongly Coupled Magnetic Resonance,\textquotedblright~\textit{IEEE Transactions on Antennas and Propagation}, vol. 61, no. 3, pp. 1378-1384, March 2013.
	
	\bibitem{Hertz-1}
	H. Hertz, \textit{Dictionary of Scientific Biography}~vol. VI. New York: Scribner, pp. 340-349.
	
	\bibitem{Brown-1}
	W. C. Brown, \textquotedblleft The History of Power Transmission by Radio Waves,\textquotedblright~\textit{IEEE Transactions on Microwave Theory and Techniques}, vol. 32, no. 9, pp. 1230-1242, September 1984.
	
	\bibitem{Tesla-1}
	N. Tesla, \textquotedblleft Apparatus for transmitting electrical energy,\textquotedblright U.S. Patent 1 119 732, December 1914.
	
	\bibitem{Tesla-2}
	N. Tesla, \textquotedblleft The true wireless,\textquotedblright~\textit{Electrical Experimenter}, May 1919.
	
	\bibitem{Shinohara-1}
	N. Shinohara, \textit{Wireless Power Transfer via Radiowaves}, London, UK: Wiley, 2014, ch. 1, pp. 1-12.
	
	\bibitem{Belo-1}
	D. Belo, D. C. Ribeiro, P. Pinho, and N. B. Carvalho, \textquotedblleft A Selective, Tracking, and Power Adaptive Far-Field Wireless Power Transfer System,\textquotedblright~\textit{IEEE Transactions on Microwave Theory and Techniques}, vol. 67, no. 9, pp. 3856-3866, September 2019.
	
	\bibitem{Garnica-1}
	J. Garnica, R. A. Chinga, and J. Lin, \textquotedblleft Wireless Power Transmission: From Far Field to Near Field,\textquotedblright~\textit{Proceedings of the IEEE}, vol. 101, no. 6, pp. 1321-1331, June 2013.
	
	\bibitem{Kung-1}
	M. Kung and K. Lin, \textquotedblleft Enhanced Analysis and Design Method of Dual-Band Coil Module for Near-Field Wireless Power Transfer Systems,\textquotedblright~\textit{IEEE Transactions on Microwave Theory and Techniques}, vol. 63, no. 3, pp. 821-832, March 2015.
	
	\bibitem{Spadden-1}
	J. O. McSpadden and J. C. Mankins, \textquotedblleft Space solar power programs and microwave wireless power transmission technology,\textquotedblright~\textit{IEEE Microwave Magazine}, vol. 3, no. 4, pp. 46-57, December 2002.
	
	\bibitem{Xu-1}
	J. Xu, Y. Zeng, and R. Zhang, \textquotedblleft UAV-Enabled Wireless Power Transfer: Trajectory Design and Energy Optimization,\textquotedblright~\textit{IEEE Transactions on Wireless Communications}, vol. 17, no. 8, pp. 5092-5106, August 2018.
	
	\bibitem{Ghotbi-1}
	I. Ghotbi, M. Najjarzadegan, H. Sarfaraz, S. J. Ashtiani, and O. Shoaei, \textquotedblleft Enhanced Power-Delivered-to-Load Through Planar Multiple-Harmonic Wireless Power Transmission,\textquotedblright~\textit{IEEE Transactions on Circuits and Systems II: Express Briefs}, vol. 65, no. 9, pp. 1219-1223, September 2018.
	
	\bibitem{Kiani-1}
	M. Kiani and M. Ghovanloo, \textquotedblleft An RFID-Based Closed-Loop Wireless Power Transmission System for Biomedical Applications,\textquotedblright~\textit{IEEE Transactions on Circuits and Systems II: Express Briefs}, vol. 57, no. 4, pp. 260-264, April 2010.
	
	\bibitem{Zargham-1}
	M. Zargham and P. G. Gulak, \textquotedblleft Fully Integrated On-Chip Coil in 0.13 $\mu {\rm m}$ CMOS for Wireless Power Transfer Through Biological Media,\textquotedblright~\textit{IEEE Transactions on Biomedical Circuits and Systems}, vol. 9, no. 2, pp. 259-271, April 2015.
	
	\bibitem{Wang-2}
	G. Wang, P. Wang, Y. Tang, and W. Liu, \textquotedblleft Analysis of Dual Band Power and Data Telemetry for Biomedical Implants,\textquotedblright~\textit{IEEE Transactions on Biomedical Circuits and Systems}, vol. 6, no. 3, pp. 208-215, June 2012.
	
	\bibitem{Chen-1}
	C. Chen, T. Chu, C. Lin, and Z. Jou, \textquotedblleft A Study of Loosely Coupled Coils for Wireless Power Transfer,\textquotedblright~\textit{IEEE Transactions on Circuits and Systems II: Express Briefs}, vol. 57, no. 7, pp. 536-540, July 2010.
	
	\bibitem{Tahar-1}
	F. Tahar, A. Barakat, R. Saad, K. Yoshitomi, and R. K. Pokharel, \textquotedblleft Dual-Band Defected Ground Structures Wireless Power Transfer System With Independent External and Inter-Resonator Coupling,\textquotedblright~\textit{IEEE Transactions on Circuits and Systems II: Express Briefs}, vol. 64, no. 12, pp. 1372-1376, December 2017.
	
	\bibitem{Cannon-1}
	B. L. Cannon, J. F. Hoburg, D. D. Stancil, and S. C. Goldstein, \textquotedblleft Magnetic Resonant Coupling As a Potential Means for Wireless Power Transfer to Multiple Small Receivers,\textquotedblright~\textit{IEEE Transactions on Power Electronics}, vol. 24, no. 7, pp. 1819-1825, July 2009.
	
	\bibitem{Jolani-2}
	F. Jolani, Y. Yu, and Z. Chen, \textquotedblleft Enhanced planar wireless power transfer using strongly coupled magnetic resonance,\textquotedblright~\textit{Electronics Letters}, vol. 51, no. 2, pp. 173-175, January 2015.
	
	\bibitem{Jolani-1}
	F. Jolani, Y. Yu, and Z. Chen, \textquotedblleft A Planar Magnetically Coupled Resonant Wireless Power Transfer System Using Printed Spiral Coils,\textquotedblright~\textit{IEEE Antennas and Wireless Propagation Letters}, vol. 13, pp. 1648-1651, August 2014.
	
	\bibitem{Mohamed-1}
	M. Mohamed, M. Cheffena, A. Moldsvor, and F. P. Fontan, \textquotedblleft Physical-Statistical Channel Model for Off-Body Area Network,\textquotedblright~\textit{IEEE Antennas and Wireless Propagation Letters}, vol. 16, pp. 1516-1519, January 2017.
	
	\bibitem{Rano-2}
	D. Rano and M. Hashmi, \textquotedblleft Extremely compact EBG-backed antenna for smartwatch applications in medical body area network,\textquotedblright~\textit{IET Microwaves, Antennas \& Propagation}, vol. 13, no. 7, pp. 1031-1040, June 2019.
	
	\bibitem{Ling-1}
	Z. Ling, F. Hu, L. Wang, J. Yu, and X. Liu, \textquotedblleft Point-to-Point Wireless Information and Power Transfer in WBAN With Energy Harvesting,\textquotedblright~\textit{IEEE Access, vol. 5}, pp. 8620-8628, April 2017.
	
	\bibitem{Huh-1}
	J. Huh, S. W. Lee, W. Y. Lee, G. H. Cho, and C. T. Rim, \textquotedblleft Narrow-Width Inductive Power Transfer System for Online Electrical Vehicles,\textquotedblright~\textit{IEEE Transactions on Power Electronics}, vol. 26, no. 12, pp. 3666-3679, December 2011.
	
	\bibitem{Hsieh-1}
	Y. Hsieh, Z. Lin, M. Chen, H. Hsieh, Y. Liu, and H. Chiu, \textquotedblleft High-Efficiency Wireless Power Transfer System for Electric Vehicle Applications,\textquotedblright~\textit{IEEE Transactions on Circuits and Systems II: Express Briefs}, vol. 64, no. 8, pp. 942-946, August 2017.
	
	\bibitem{Jia-1}
	Y. Jia, A. Mirbozorgi, Z. Wang, C. Hsu, T. E. Madsen,  D. Rainnie, and M. Ghovanloo \textquotedblleft Position and Orientation Insensitive Wireless Power Transmission for EnerCage-Homecage System,\textquotedblright~\textit{IEEE Transactions on Biomedical Engineering}, vol. 64, no. 10, pp. 2439-2449, October 2017.
	
	\bibitem{Mei-1}
	H. Mei, K. A. Thackston, R. A. Bercich, J. G. R. Jefferys, and P. P. Irazoqui, \textquotedblleft Cavity Resonator Wireless Power Transfer System for Freely Moving Animal Experiments,\textquotedblright~\textit{IEEE Transactions on Biomedical Engineering}, vol. 64, no. 4, pp. 775-785, April 2017.
	
	\bibitem{Freeman-1}
	D. K. Freeman and S. J. Byrnes, \textquotedblleft Optimal Frequency for Wireless Power Transmission Into the Body: Efficiency Versus Received Power,\textquotedblright~\textit{IEEE Transactions on Antennas and Propagation}, vol. 67, no. 6, pp. 4073-4083, June 2019.
	
	\bibitem{Jow-1}
	U. Jow and M. Ghovanloo, \textquotedblleft Design and Optimization of Printed Spiral Coils for Efficient Transcutaneous Inductive Power Transmission,\textquotedblright~\textit{IEEE Transactions on Biomedical Circuits and Systems}, vol. 1, no. 3, pp. 193-202, September 2007.
	
	\bibitem{Zhang-1}
	Y. Zhang, Z. Zhao, and K. Chen, \textquotedblleft Frequency-Splitting Analysis of Four-Coil Resonant Wireless Power Transfer,\textquotedblright~\textit{IEEE Transactions on Industry Applications}, vol. 50, no. 4, pp. 2436-2445, August 2014.
	
	\bibitem{R1}
	S. Yu, H. Liu, and L. Li, \textquotedblleft Design of Near-Field Focused Metasurface for High-Efficient Wireless Power Transfer With Multifocus Characteristics,\textquotedblright~\textit{IEEE Transactions on Industrial Electronics}, vol. 66, no. 5, pp. 3993-4002, May 2019.
	
	\bibitem{R2}
	P. Zhang, L. Li, X. Zhang, H. Liu, and Y. Shi, \textquotedblleft Design, Measurement and Analysis of Near-Field Focusing Reflective Metasurface for Dual-Polarization and Multi-Focus Wireless Power Transfer,\textquotedblright~\textit{IEEE Access}, vol. 7, pp. 110387-110399, August 2019.
	
	\bibitem{Li-1}
	L. Li, H. Liu, H. Zhang, and W. Xue, \textquotedblleft Efficient Wireless Power Transfer System Integrating With Metasurface for Biological Applications,\textquotedblright~\textit{IEEE Transactions on Industrial Electronics}, vol. 65, no. 4, pp. 3230-3239, April 2018.
	
	\bibitem{Liu-3}
	C. Liu, A. P. Hu, and N. K. C. Nair, \textquotedblleft Modelling and analysis of a capacitively coupled contactless power transfer system,\textquotedblright~\textit{IET Power Electronics}, vol. 4, no. 7, pp. 808-815, August 2011.
	
	\bibitem{R3}
	L. Li, X. Zhang, C. Song, and Y. Huang, \textquotedblleft Progress, challenges, and perspective on metasurfaces for ambient radio frequency energy harvesting,\textquotedblright~\textit{Applied Physics Letters}, 116, 060501, 2020.
	
	\bibitem{Liou-1}
	C. Liou, C. Kuo, and S. Mao, \textquotedblleft Wireless-Power-Transfer System Using Near-Field Capacitively Coupled Resonators,\textquotedblright~\textit{IEEE Transactions on Circuits and Systems II: Express Briefs}, vol. 63, no. 9, pp. 898-902, September 2016.
	
	\bibitem{Riehl-1}
	P. S. Riehl, A. Satyamoorthy, H. Akram, Y. Yen, J. C. Yang, B. Juan, C. Lee, F. Lin, V. Muratov, W. Plumb, and P. F. Tustin, \textquotedblleft Wireless Power Systems for Mobile Devices Supporting Inductive and Resonant Operating Modes,\textquotedblright~\textit{IEEE Transactions on Microwave Theory and Techniques}, vol. 63, no. 3, pp. 780-790, March 2015.
	
	\bibitem{Kurs-1}
	A. Kurs, A. Karalis, R. Moffat, J. D. Joannopoulos, P. Fisher, and M. Solja{\v c}i{\'c}, \textquotedblleft Wireless Power Transfer via Strongly Coupled Magnetic Resonances,\textquotedblright~\textit{Science},vol. 317, no. 5834, pp. 83-85, July 2007.
	
	\bibitem{Lu-1}
	X. Lu, P. Wang, D. Niyato, D. I. Kim, and Z. Han, \textquotedblleft Wireless Charging Technologies: Fundamentals, Standards, and Network Applications,\textquotedblright~\textit{IEEE Communications Surveys \& Tutorials}, vol. 18, no. 2, pp. 1413-1452, November 2015.
	
	\bibitem{Kim-1}
	H. Kim, H. Hirayama, S. Kim, K. J. Han, R. Zhang, and J. Choi, \textquotedblleft Review of Near-Field Wireless Power and Communication for Biomedical Applications,\textquotedblright~\textit{IEEE Access}, vol. 5, pp. 21264-21285, September 2017.
	
	\bibitem{Wang-1}
	J. Wang, J. Li, S. L. Ho, W. Y, Chau, W. K. Lee, W. N. Fu, Y. Li, H. Yu, and M. Sun, \textquotedblleft Study and Experimental Verification of a Rectangular Printed-Circuit-Board Wireless Transfer System for Low Power Devices,\textquotedblright~\textit{IEEE Transactions on Magnetics}, vol. 48, no. 11, pp. 3013-3016, November 2012.
	
	\bibitem{Falavarjani-1}
	M. M. Falavarjani, M. Shahabadi, and J. Rashed-Mohassel, \textquotedblleft Design and implementation of compact WPT system using printed spiral resonators,\textquotedblright~\textit{Electronics Letters}, vol. 50, no. 2, pp. 110-111, January 2014.
	
	\bibitem{Barman-1}
	S. D. Barman, A. W. Reza, N. Kumar, Md. E. Karim, and A. B. Munir, \textquotedblleft Wireless powering by magnetic resonant coupling: Recent trends in wireless power transfer system and its applications,\textquotedblright~\textit{Renewable and Sustainable Energy Reviews}, vol. 51, pp. 1525-1552, July 2015.
	
	\bibitem{Hui-1}
	S. Y. R. Hui, W. Zhong, and C. K. Lee, \textquotedblleft A Critical Review of Recent Progress in Mid-Range Wireless Power Transfer,\textquotedblright~\textit{IEEE Transactions on Power Electronics}, vol. 29, no. 9, pp. 4500-4511, September 2014.
	
	\bibitem{Shinohara-2}
	N. Shinohara, \textquotedblleft Power without wires,\textquotedblright~\textit{IEEE Microwave Magazine}, vol. 12, no. 7, pp. S64-S73, December 2011.
	
	\bibitem{Hekal-2}
	S. Hekal and A. B. Abdel-Rahman, \textquotedblleft New compact design for short range wireless power transmission at 1GHz using H-slot resonators,\textquotedblright~\textit{9th European Conference on Antennas and Propagation (EuCAP)}, Lisbon, Portugal, pp. 1-5, April 2015.
	
	\bibitem{Hekal-4}
	S. Hekal, A. B. Abdel-Rahman, H. Jia, A. Allam, A. Barakat, T. Kaho, and R. K. Pokharel, \textquotedblleft Compact Wireless Power Transfer System Using Defected Ground Bandstop Filters,\textquotedblright~\textit{IEEE Microwave and Wireless Components Letters}, vol. 26, no. 10, pp. 849-851, October 2016.
	
	\bibitem{Liu-2}
	D. Liu, H. Hu, and S. V. Georgakopoulos, \textquotedblleft Misalignment Sensitivity of Strongly Coupled Wireless Power Transfer Systems,\textquotedblright~\textit{IEEE Transactions on Power Electronics}, vol. 32, no. 7, pp. 5509-5519, July 2017.
	
	\bibitem{Lee-1}
	J. Lee, Y. Lim, W. Yang, and S. Lim, \textquotedblleft Wireless Power Transfer System Adaptive to Change in Coil Separation,\textquotedblright~\textit{IEEE Transactions on Antennas and Propagation}, vol. 62, no. 2, pp. 889-897, February 2014.
	
	\bibitem{Raju-1}
	S. Raju, R. Wu, M. Chan, and C. P. Yue, \textquotedblleft Modeling of Mutual Coupling Between Planar Inductors in Wireless Power Applications,\textquotedblright~\textit{IEEE Transactions on Power Electronics}, vol. 29, no. 1, pp. 481-490, January 2014.
	
	\bibitem{Mastri-1}
	F. Mastri, A. Costanzo, and M. Mongiardo, \textquotedblleft Coupling-Independent Wireless Power Transfer,\textquotedblright~\textit{IEEE Microwave and Wireless Components Letters}, vol. 26, no. 3, pp. 222-224, March 2016.
	
	\bibitem{Li-2}
	L. Li, K. Ma, and S. Mou, \textquotedblleft Modeling of New Spiral Inductor Based on Substrate Integrated Suspended Line Technology,\textquotedblright~\textit{IEEE Transactions on Microwave Theory and Techniques}, vol. 65, no. 8, pp. 2672-2680, August 2017.
	
	\bibitem{Ahn-2}
	D. Ahn and P. P. Mercier, \textquotedblleft Wireless Power Transfer With Concurrent 200-kHz and 6.78-MHz Operation in a Single-Transmitter Device,\textquotedblright~\textit{IEEE Transactions on Power Electronics}, vol. 31, no. 7, pp. 5018-5029, July 2016.
	
	\bibitem{Ghovanloo-1}
	M. Ghovanloo and S. Atluri, \textquotedblleft A Wide-Band Power-Efficient Inductive Wireless Link for Implantable Microelectronic Devices Using Multiple Carriers,\textquotedblright~\textit{IEEE Transactions on Circuits and Systems I: Regular Papers}, vol. 54, no. 10, pp. 2211-2221, October 2007.
	
	\bibitem{Hekal-1}
	S. Hekal, A. B. Abdel-Rahman, H. Jia, A. Allam, A. Barakat, and R. K. Pokharel, \textquotedblleft A Novel Technique for Compact Size Wireless Power Transfer Applications Using Defected Ground Structures,\textquotedblright~\textit{IEEE Transactions on Microwave Theory and Techniques}, vol. 65, no. 2, pp. 591-599, February 2017.
	
	\bibitem{Lim-7}
	J. Lim, B. Lee, and M. Ghovanloo, \textquotedblleft Optimal Design of a Resonance-Based Voltage Boosting Rectifier for Wireless Power Transmission,\textquotedblright~\textit{IEEE Transactions on Industrial Electronics}, vol. 65, no. 2, pp. 1645-1654, February 2018.
	
	\bibitem{RamRakhyani-1}
	A. K. RamRakhyani and G. Lazzi, \textquotedblleft On the Design of Efficient Multi-Coil Telemetry System for Biomedical Implants,\textquotedblright~\textit{IEEE Transactions on Biomedical Circuits and Systems}, vol. 7, no. 1, pp. 11-23, February 2013.
	
	\bibitem{Ahn-3}
	D. Ahn and M. Ghovanloo, \textquotedblleft Optimal Design of Wireless Power Transmission Links for Millimeter-Sized Biomedical Implants,\textquotedblright~\textit{IEEE Transactions on Biomedical Circuits and Systems}, vol. 10, no. 1, pp. 125-137, February 2016.
	
	\bibitem{Lou-1}
	X. Lou and G. Yang, \textquotedblleft A Dual Linearly Polarized Rectenna Using Defected Ground Structure for Wireless Power Transmission,\textquotedblright~\textit{IEEE Microwave and Wireless Components Letters}, vol. 28, no. 9, pp. 828-830, September 2018.
	
	\bibitem{Verma-1}
	S. Verma, D. Rano, M. Hashmi, and V. Bohara, \textquotedblleft A High Q Dual E-Shaped Defected Ground Structure for Wireless Power Transfer Applications,\textquotedblright~\textit{Asia-Pacific Microwave Conference (APMC)}, Kyoto, Japan, pp. 1435-1437, November 2018.
	
	\bibitem{Chalise-1}
	S. Chalise, F. Tahar, M. R. Saad, A. Baraket, K. Yoshitomi, and R. K. Pokharel, \textquotedblleft High Efficiency Wireless Power Transfer System Using Spiral DGS Resonators Through Biological Tissues,\textquotedblright~\textit{IEEE International Microwave Biomedical Conference (IMBioC)}, Philadelphia, PA, pp. 43-45, June 2018.
	
	\bibitem{My-2}
	K. Dautov, R. Gupta, and M. Hashmi, \textquotedblleft A Performance Enhanced Dual-band Wireless Power Transfer System for Practical ISM Bands,\textquotedblright~\textit{Asia-Pacific Microwave Conference (APMC)}, Singapore, pp. 1259-1261, December 2019.
	
	\bibitem{Verma-2}
	S. Verma, D. Rano, and M. Hashmi, \textquotedblleft A Novel Dual Band Defected Ground Structure for Short Range Wireless Power Transfer Applications,\textquotedblright~\textit{IEEE Wireless Power Week (WPW)}, London, UK, pp. 1-4, June 2019.
	
	\bibitem{Karmakar-1}
	N. C. Karmakar, S. M. Roy, and I. Balbin, \textquotedblleft Quasi-static modeling of defected ground structure,\textquotedblright~\textit{IEEE Transactions on Microwave Theory and Techniques}, vol. 54, no. 5, pp. 2160-2168, May 2006.
	
	\bibitem{Hekal-5}
	S. Hekal, A. B. Abdel-Rahman, A. Allam, A. Barakat, H. Jia, and R. K. Pokharel, \textquotedblleft Asymmetric strongly coupled printed resonators for wireless charging applications,\textquotedblright~\textit{IEEE 17th Annual Wireless and Microwave Technology Conference (WAMICON)}, Clearwater, FL, pp. 1-5, April 2016.
	
	\bibitem{Kupreyev-1}
	D. Kupreyev, K. Dautov, M. Hashmi, and S. Verma, \textquotedblleft Design of a Compact DGS based Dual-band RF WPT System for Low-Power Applications,\textquotedblright~\textit{IEEE Asia-Pacific Conference on Antennas and Propagation}, Incheon, Korea, pp. 1-2, August 2019.
	
	\bibitem{Barakat-1}
	A. Barakat, K. Yoshitomi, and R. K. Pokharel, \textquotedblleft Design Approach for Efficient Wireless Power Transfer Systems During Lateral Misalignment,\textquotedblright~\textit{IEEE Transactions on Microwave Theory and Techniques}, vol. 66, no. 9, pp. 4170-4177, September 2018.
	
	\bibitem{Barakat-3}
	A. Barakat, S. Alshhawy, K. Yoshitomi, and R. K. Pokharel, \textquotedblleft Triple-Band Near-Field Wireless Power Transfer System Using Coupled Defected Ground Structure Band Stop Filters,\textquotedblright~\textit{IEEE MTT-S International Microwave Symposium (IMS)}, Boston, MA, USA, pp. 1411-1414, June 2019.
	
	\bibitem{Tahar-2}
	F. Tahar, R. Saad, A. Barakat, and R. K. Pokharel, \textquotedblleft 1.06 FoM and Compact Wireless Power Transfer System Using Rectangular Defected Ground Structure Resonators,\textquotedblright~\textit{IEEE Microwave and Wireless Components Letters}, vol. 27, no. 11, pp. 1025-1027, November 2017.
	
	\bibitem{My-1}
	K. Dautov, M. Hashmi, S. Verma, and R. Gupta, \textquotedblleft Highly-efficient Compact Size DGS-based Wireless Power Transfer for Low-Power Sensor Nodes,\textquotedblright~\textit{IEEE MTT-S International Microwave and RF Conference (IMaRC)}, Mumbai, India, pp. 1-3, December 2019.
	
	\bibitem{Barakat-4}
	A. Barakat, K. Yoshitomi, and R. K. Pokharel, \textquotedblleft Design and Implementation of Dual-Mode Inductors for Dual-Band Wireless Power Transfer Systems,\textquotedblright~\textit{IEEE Transactions on Circuits and Systems II: Express Briefs}, vol. 66, no. 8, pp. 1287-1291, August 2019.
	
	\bibitem{Sharaf-2}
	R. Sharaf, A. B. Abdel-Rahman, A. S. Abd El-Hameed, A. Barakat, S. Hekal, and A. Allam, \textquotedblleft A New Compact Dual-Band Wireless Power Transfer System Using Interlaced Resonators,\textquotedblright~\textit{IEEE Microwave and Wireless Components Letters}, vol. 29, no. 7, pp. 498-500, July 2019.
	
	\bibitem{Lim}
	J. Lim, Y. Lee, C. Kim, D. Ahn, and S. Nam, \textquotedblleft A vertically periodic defected ground structure and its application in reducing the size of microwave circuits,\textquotedblright~\textit{IEEE Microwave and Wireless Components Letters}, vol. 12, no. 12, pp. 479-481, December 2002.
	
	\bibitem{Huang-1}
	S. Y. Huang and Y. H. Lee, \textquotedblleft A Compact E-Shaped Patterned Ground Structure and Its Applications to Tunable Bandstop Resonator,\textquotedblright~\textit{IEEE Transactions on Microwave Theory and Techniques}, vol. 57, no. 3, pp. 657-666, March 2009.
	
	\bibitem{Kim-2}
	C. Kim, J. Park, D. Ahn, and J. Lim, \textquotedblleft A novel 1-D periodic defected ground structure for planar circuits,\textquotedblright~\textit{IEEE Microwave and Guided Wave Letters}, vol. 10, no. 4, pp. 131-133, April 2000.
	
	\bibitem{Ahn-1}
	D. Ahn, J.Park, C. Kim, J. Kim, Y. Qian, and T. Itoh, \textquotedblleft A design of the low-pass filter using the novel microstrip defected ground structure,\textquotedblright~\textit{IEEE Transactions on Microwave Theory and Techniques}, vol. 49, no. 1, pp. 86-93, January 2001.
	
	\bibitem{Mandal-1}
	M. K. Mandal and S. Sanyal, \textquotedblleft A novel defected ground structure for planar circuits,\textquotedblright~\textit{IEEE Microwave and Wireless Components Letters}, vol. 16, no. 2, pp. 93-95, February 2006.
	
	\bibitem{Abdel-Rahman-1}
	A. Abdel-Rahman, A. K. Verma, A. Boutejdar, and A. S. Omar, \textquotedblleft Compact stub type microstrip bandpass filter using defected ground plane,\textquotedblright~\textit{IEEE Microwave and Wireless Components Letters}, vol. 14, no. 4, pp. 136-138, April 2004.
	
	\bibitem{Liu-1}
	H. W. Liu, Z. F. Li, X. W. Sun, and J. F. Mao, \textquotedblleft An improved 1D periodic defected ground structure for microstrip line,\textquotedblright~\textit{IEEE Microwave and Wireless Components Letters}, vol. 14, no. 4, pp. 180-182, April 2004.
	
	\bibitem{Woo}
	D. Woo, T. Lee, J.Lee, C. Pyo, and W. Choi, \textquotedblleft Novel U-slot and V-slot DGSs for bandstop filter with improved $Q$ factor,\textquotedblright~\textit{IEEE Transactions on Microwave Theory and Techniques}, vol. 54, no. 6, pp. 2840-2847, June 2006.
	
	\bibitem{Liu-4}
	H. Liu, Z. Li, and X. Sun, \textquotedblleft Compact defected ground structure in microstrip technology,\textquotedblright~\textit{Electronics Letters}, vol. 41, no. 3, pp. 132-134, 3 February 2005.
	
	\bibitem{Wang}
	X. Wang, B. Wang, H. Zhang, and K. J. Chen, \textquotedblleft A Tunable Bandstop Resonator Based on a Compact Slotted Ground Structure, \textquotedblright~\textit{IEEE Transactions on Microwave Theory and Techniques}, vol. 55, no. 9, pp. 1912-1918, September 2007.
	
	\bibitem{Kumar}
	C. Kumar and D. Guha, \textquotedblleft Asymmetric Geometry of Defected Ground Structure for Rectangular Microstrip: A New Approach to Reduce its Cross-Polarized Fields,\textquotedblright~\textit{IEEE Transactions on Antennas and Propagation}, vol. 64, no. 6, pp. 2503-2506, June 2016.
	
	\bibitem{Imura-1}
	T. Imura and Y. Hori, \textquotedblleft Maximizing Air Gap and Efficiency of Magnetic Resonant Coupling for Wireless Power Transfer Using Equivalent Circuit and Neumann Formula,\textquotedblright~\textit{IEEE Transactions on Industrial Electronics}, vol. 58, no. 10, pp. 4746-4752, October 2011.
	
	
	
	
	
	
	
	
	
	
	
	
	
	
	
	
	
\end{thebibliography}
\end{document}